# Statements of Current States of the Art for Key non-Coronagraphic Technologies for HWO

Paul Scowen, HWO UVI Instrument Scientist, NASA Goddard, paul.a.scowen@nasa.gov
Manuel Quijada, NASA Goddard, manuel.a.quijada@nasa.gov
Emily Kan, NASA Goddard, emily.kan@nasa.gov
Michael Hoenk, NASA JPL, michael.e.hoenk@jpl.nasa.gov
Prabal Saxena, NASA Goddard, prabal.saxena@nasa.gov
Oswald Siegmund, Sensor Sciences, osiegmund@sensorsciences.com
Alexander Kutyrev, NASA Goddard, alexander.s.kutyrev@nasa.gov
Massimo Roberto, STScI, robberto@stsci.edu
Randy McEntaffer, Penn State Univ., rlm90@psu.edu
Juan Larruquert, CSIC Madrid, Spain, j.larruquert@csic.es

## Far-UV Mirror Coating

Proposal partnership POC: Manuel Quijada, NASA Goddard, manuel.a.quijada@nasa.gov

## Discussion

The mirror coatings necessary for HWO involve the use of materials and deposition techniques that can deliver the appropriate reflectivity down to 100nm wavelength, level of uniformity, knowledge of both polarization aberration terms, and the resilience and stability for a long mission lifetime at the conditions typical of L2.

Note that the performance of the mirror coatings is regarded as a threshold capability since it enables both the wavefront control necessary for the coronagraphy and the throughput down to 100nm needed for transformational astrophysics prioritized by the Decadal Survey 2020.

**Current State of the Art**
The current Technology Readiness Level of the current best candidate for FUV mirror coating performance, Al+XeLiF, has been demonstrated to be at the 3 level (see Quijada *et al.* (2026); Quijada *et al.* (2025); Lewis *et al.* (2024); Quijada *et al.* (2024); Quijada *et al.* (2022)).
- Analytical and Experimental Proof-of-Concept:
    - Theoretical models predicted the expected behavior of the Al+XeLiF reflectance and polarization properties
    - Experimental validation of these predictions demonstrated on substrates of glass, ULE and Zerodur.
- Performance Parameters:
    - Properties such as spectral performance, polarization, roughness, durability, and environmental stability have been characterized (see references).
    - Preliminary testing demonstrate a state-of-the-art reflectance performance in the far-Ultraviolet (FUV) spectral range with R over 90% at H-Lyman-Alpha wavelength (121.6nm) and over 70% at H-Lyman Beta wavelength (102.6 nm), while preserving the high intrinsic reflectivity of bare Al in the NUV/Vis/IR. Preliminary testing shows coating surface roughness (RMS) below 1 nm.

> Preliminary polarization studies match the predictions. Preliminary testing indicates that the coating show little degradation over large periods of storage, and after environmental conditions of room-temperature (25°C) and up to 60% relative humidity. Adhesion tests are pending.

- Initial Prototypes :
  - Coatings are deposited on small (1-2 inch) substrates or using laboratory equipment.
  - Results show scaling up to larger substrates is feasible.
- Measured Properties and Application Requirements:
  - Measurements show a state-of-the-art reflectance performance in the far-Ultraviolet (FUV) spectral range with R over 90% at H-Lyman-Alpha wavelength (121.6nm). Optical properties have been compared against HWO, LUVOIR, and HabEx spectral targets.
  - This coating has been deployed in two different flight missions: Two gratings have been coated with Al+XeLiF for the INFUSE sounding rocket and the SPRITE CubeSAT projects (PI: Brian Fleming U. of CO).
- Repeatability and Reproducibility:
  - The fabrication process is repeatable and reproducible, with consistent results across multiple coated samples, as indicated during the MISSE-20 experiments, and samples from different coating runs exhibit similar and consistent spectral performance.

# Near UV/VIS Detectors

## Discussion

The provision of detectors that perform in the near ultraviolet (NUV) and visible wavelength bands is a baseline capability necessary to enable both the high resolution imager (HRI) instrument and the ultraviolet instrument (UVI, both MOS and IFU modes) instrument. While these capabilities are also needed for the various coronagraphic channels in Section 4 above, this discussion will purely focus on the needs of the astrophysics instruments and what needs to be done to deliver their capabilities.

## Current State of the Art

### Thick, p-channel MAS-CCD

Proposal partnership POC: Emily Kan, NASA Goddard, emily.kan@nasa.gov

The MAS-CCD inherits from the thick, fully-depleted p-channel architecture and non-destructive measurement capabilities of the Skipper CCDs developed for particle physics applications (Holland (2023)). The MAS-CCD implements multiple floating gate Skipper amplifiers on the CCD serial register to perform multiple, non-destructive measurements of the charge in each detector pixel. In contrast to the conventional Skipper CCD architecture (Tiffenberg et al (2017)), which performs multiple measurements sequentially with a single amplifier, the MAS architecture can combine non-destructive measurements from multiple amplifiers. This results in a reduction of readout noise proportional to the square-root of the number of amplifiers with a minimal (~10%) increase in readout time relative to the single-amplifier, single-measurements conventional CCD architecture (Holland (2023); Tiffenberg et al (2017); Botti et al (2024); Lapi et al (2024); Lin et al (2024)).

*Thick, p-channel SiSeRO CCD*

Proposal partnership POC: Emily Kan, NASA Goddard, emily.kan@nasa.gov

Skipper-CCDs have emerged as a powerful tool in astrophysics and astronomy, demonstrating sensitivity to single electrons and single photons. Thick p-channel, skipper-CCD sensors present a unique opportunity for photon-counting detectors with sensitivity from the near-UV to the near-IR. Despite their recent success, the current readout speed of Skipper-CCDs limits their practical applications. To enhance their usability for space applications, including HWO, improvements in readout speed are necessary. One potential solution is the MAS-CCD, as mentioned previously, which aims to achieve faster readout. The SiSeRO amplifier offers another innovative approach by introducing a new method for sensing pixel charge in CCDs.

The FGA in the skipper-CCD works by transferring the pixel charge onto the bulk side of a MOS capacitor in the CCD buried channel to regulate the gate voltage of an output transistor. However, in addition to the MOS capacitance, various stray capacitances are added to the floating gate, which limits the sensitivity of the FGA stage. The overall FGA capacitance, along with the resistance of the large polysilicon connection between the MOS capacitor and the output transistor gate, restricts the time response of the FGA (i.e., slowing it down). As a result, the FGA maximum pixel readout rate is limited. The limited sensitivity of the FGA combined with the multisampling operations significantly increases the CCD readout time.

To address the readout noise and speed limitations of FGAs, the SiSeRO amplifier incorporates a double-gate MOSFET that is integrated into the buried channel of the CCD. In this configuration, the CCD channel is junction-coupled to the MOSFET channel, resulting in high-sensitivity modulation of the MOSFET current by the signal charge in the CCD channel. The SiSeRO output stage is specifically designed to be compatible with LBNL's fully-depleted p-channel CCDs to allow for sensitivity from the near-UV to the near-IR (Sofo-Haro(2023)). Currently, the SiSeRO amplifier is being evaluated for the readout of modern n-channel thin CCDs (Chattopadhyay(2022)). In addition to its high sensitivity, another advantage of this amplifier is its ability to perform RNDR (repetitive non-destructive readout). With higher sensitivity to the pixel charge, SiSeRO-CCDs provide sub-electron noise and single-photon counting capabilities in shorter readout times than Skipper-CCDs.

*EMCCD*

Proposal partnership POC: Michael Hoenk, NASA JPL, michael.e.hoenk@jpl.nasa.gov

Teledyne-e2v's CCD311 is a radiation hardened version of their EMCCD technology. The CCD311 was delivered for the Coronagraph Instrument on the Roman Space Telescope (Roman CGI; Bush et al 2025; Morrissey et al 2023), and current TRL for that mission is TRL 8. If detector performance requirements for the HWO coronagraph instrument are assumed to be similar to that of Roman CGI except for the mission duration (10 yr. vs 5 yr.), then TRL is assessed to be 4+ for HWO and a TRL advancement program is needed to qualify the CCD311 for the HWO radiation environment.

| Parameter | "Peak Demonstrated Performance" | Roman CGI | Peak performance capability OK for HWO? | Notes |
|---|---|---|---|---|
| Read Noise ($e^-$) | <0.02 | <0.02 | Yes | Noise w/ EM gain applied. Reduction in base read noise beneficial for HWO class implementation. Final Roman CGI read noise – 165 $e^-$ at 10 MHz. ABB/Nuvu Camera capability – 100 $e^-$. Normal CCD output (minimum 8 $e^-$ at 1 MHz) should be used for observations that do not require EM gain. |
| Dark Current ($e^-$/pix/s) | $< 1.5 \times 10^{-5}$ | $1.5 \times 10^{-5}$ | Yes | Dark current with radiation dose consistent with 10 years needs to be measured with cryogenic irradiation. Effect of annealing needs to be measured. |
| Spurious Noise (e-/pix/frame) | $6.9 \times 10^{-4}$ | $8.6 \times 10^{-3}$ | Yes | Measured with applied EM gain equal to 50 × read noise, such that DQE = 90 %. |
| Format | 4k × 4k | 2k × 1k | Borderline | Desire is for 2k × 2k sensor per Exoplanet Technology Gap Study Report. https://exoplanets.nasa.gov/internal_resources/2595/ |
| Detector Lifetime | 5 years | 5 years | No | Need to demonstrate the sensor meets requirements when subjected to a radiation dose consistent with 10 years in orbit. Ref/ SAT submitted 2024 and to be submitted for 2025. |

*UV CMOS*

Proposal partnership POC: Michael Hoenk, NASA JPL, michael.e.hoenk@jpl.nasa.gov

The earliest manuscript describing delta-doped CMOS detectors was published in 2009 (Hoenk et al (2009). The next several years focused on demonstrating UV sensitivity optimization and performance stability (Hoenk et al (2013); Hoenk et al (2014); Hoenk et al (2015)). Recent work is focused on demonstrating wafer-scale delta-doping to produce large-format UV/Visible detectors, including 2k x 2k (Cote et al (in prep)), 4k x 2k (Greffe et al (2022)) as well as 4k x 4k and 8k x 8k (Jewell et al (2024)). Also, the Ultraviolet Explorer (UVEX) Medium Explorer (MIDEX) mission has baselined 4k x 4k delta-doped CMOS image sensor for science observations. Delta-doped silicon detectors have been flown on sounding rockets and balloon experiments, including 4k x 2k CMOS image sensors (Sanchez-Maes et al (2024)) and 1k x 2k electron multiplying CCDs (EMCCDs) (Hamden et al (2020)). Development and maturation of delta-doped CMOS image sensors to address HWO technology gaps for general astronomy and exoplanet science is underway. It is important to note that the device format—number and size of pixels, buttability— is decided by the manufacturer; customizations required for HWO require close collaboration with industry partner(s).

*NUV SiC*

Proposal partnership POC: Prabal Saxena, NASA Goddard, prabal.saxena@nasa.gov

Work on the development (Saxena et al 2025) of highly sensitive wide energy band gap ($E_g$) UV-EUV detectors based on 4H-SiC ($E_g$ = 3.2 eV) at NASA Goddard Space Flight Center (GSFC) started more than two decades ago with the aim of fabricating photodiodes which offered several key advantages for space applications over conventional, narrow bandgap, Si ($E_g$ = 1.1 eV) based devices including:
- Intrinsic carrier density is $10^{20}$ times lower than that of Si, leading to significantly lower dark current
- Visible blindness enabling UV, EUV, and X-ray photon detection without interference from visible or IR backgrounds
- High displacement energy (~ 21 eV) which offers superior radiation hardness and high breakdown electric fields
- Chemical inertness and mechanical strength offered by extremely robust silicon-carbon covalent bonds

Furthermore, the shorter cut-off wavelength of these material systems eliminates the need for bulky and expensive optical filtering components, mitigating risk, increasing system quantum efficiency (QE) and allowing for simpler optical designs of instrumentation. The most compelling reason for using wide-band gap semiconductors like 4H-SiC is the low dark currents that can be attained. SiC inherently demonstrates multiple orders of magnitude lower dark current than similar UV detecting materials like silicon by virtue of the reduced intrinsic carrier concentration even at ambient temperatures. Dark current in large bandgap 4H-SiC is also extremely responsive to temperature allowing significant reduction with minimal cooling. The dark current in the depletion region of a semiconductor, under zero or reverse bias conditions, is directly proportional to the intrinsic carrier concentration ($n_i$) given by,

$$n_i = N_v N_c \exp\left(-\frac{E_g}{2kT}\right)$$

where $N_v$ is the valence band density of states, $N_c$ is the conduction band density of states, $E_g$ is the energy band gap, k is the Boltzmann constant, and T is the temperature. This indicates that large bandgap 4H-SiC photodiodes should demonstrate a further, more than factor of 2 reduction in dark current on cooling by just an additional 4 K, allowing achievement of ultralow dark current with minimal cooling, reducing detector size weight and power consumption, SWaP-C, and preventing issues with cold trapping molecules on cryogenic detector surfaces. SiC also demonstrates the greatest material maturity of all wide-bandgap materials evaluated for production of UV photodetectors (Cho et al., 2025), for example, the defect density of SiC ($10$–$10^3$/cm$^3$) is several orders of magnitude lower than another wide bandgap semiconductor, GaN ($10^6$–$10^{10}$/cm$^3$). By virtue of these inherent benefits, and ongoing development over the last 2 decades, SiC photodiode based focal plane array technology has advanced to the point that it should be considered an incredibly promising option for HWO instruments.

Currently SiC arrays are being demonstrated with quantum efficiencies (Figure 1a), dark noise (Figure 1b), feature sizes (Figure 1c), and noise equivalent powers (Figure 1d) approaching those needed for HWO instruments including UV Spectrographs, Coronagraphs, and High Resolution Imagers. Furthermore preliminary simulations of SiC performance in UV Coronagraph instruments observing an Exo-Earths in the UV (Figure 1e) indicate SiC offers considerably reduced error in measured contrast ratio compared to measurements made with state of the art silicon detectors.

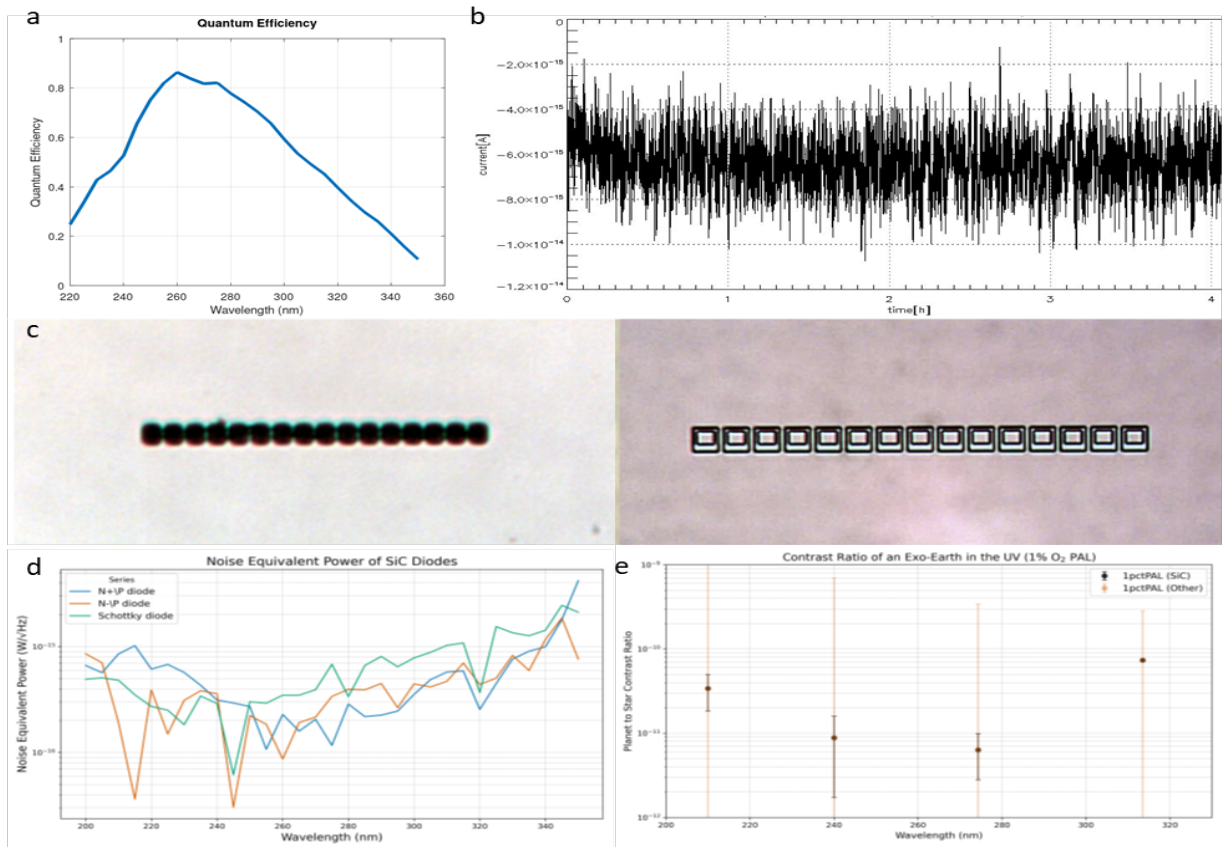

Figure 1. Recent results from SiC detector Characterization, Fabrication, and Modeling including: (a) Quantum efficiency of recently fabricated N- on P, SiC photodiodes, (b) long term collection of current data from an unilluminated SiC photodiode demonstrating a dark current ceiling of no greater the 6 fA, (c) micrographs of dopant implantation sites on a wafer which features diode arrays with 6 um, 10 um, 15 um, and 25 um pitches, (d) extracted noise equivalent power data for multiple photodiode structures, and (e) simulated error in extracted planet to star contrast ratio as a function of wavelength using SiC detectors (black) and silicon detectors (yellow).

# Far-UV Detectors

## Discussion

FUV detectors are fundamentally different technologies than the more conventional solid state detectors used in the previous section. FUV detectors are a long-standing technology that usually involves micro-channel plate (MCP) detectors, which is a 40+ year old mission proven technology. Modern variants using ALD deposited materials and borosilicate glass enable large customizable, and stable, form factors. Photocathode choices are used to tailor the spectral response. These devices are typically operated at room temperatures. Open face FUV devices that operate down to below 120nm are more sensitive to contamination than sealed window NUV detectors, but have proven stable on long term planetary missions. MCP detectors are "photon counting" asynchronous devices allowing images to be built using the photon X and Y positions and arrival time with intrinsic time resolution of less than a microsecond, they do not have any frame read time. We discuss in this section a couple of potential approaches that can be used to provide sensitivity down to 100nm that the HWO science drivers are asking for in a low-risk but high performance way.

### Current State of the Art

Proposal partnership POC: Oswald Siegmund, Sensor Sciences, osiegmund@sensorsciences.com

The current state of the art implementations of MCP detectors and potential for optimizations for HWO needs are shaped around the following demonstrated capabilities. The concept is open face and sealed tube MCP based detectors in 100 mm format covering 100 nm – 150 nm for open face and 140 – 300 nm for sealed tube, capable of tiling to >200 mm × 400 mm.

| | |
|---|---|
| Quantum Efficiency (QE) | Alkali halide >40% 100 nm – 140 nm, Bialkali ~35%, 140 nm – 200 nm |
| Out of band rejection | $> 1\times10^{-5}$ |
| Flat field modulation | Borosilicate MCPs, stable ≤20% hex pattern ~1 resel wide. |
| MCP gain stability | $>10^{14}$ events cm$^{-2}$ |
| Intrinsic background rates | $<4\times10^{-6}$ events resel$^{-1}$ sec$^{-1}$ |
| L2 background rates | Employ proven [Chandra] GCR background reduction techniques. |
| Radiation sensitivity | Borosilicate MCP with <0.7% MeV gamma efficiency. |
| Spatial resolution | <25 μm FWHM sampled at ~6 μm electronic binning |
| Local event rate | 100 events resel$^{-1}$ sec$^{-1}$ |
| Global event rate | 1 MHz at 5% deadtime, 4 MHz at 25% deadtime. |
| Environmental | No cooling, detector and electronics Rad Hard for L2 |

The TRL starting points include:
- Smaller 50 × 50 mm$^2$ sealed tube cross strip (XS) detector with HWO goal spatial resolution at TRL5/6 (per SAT review) (Siegmund et al (2023))
- Lower performance SMT electronics for 100 mm format XS at TRL6+ (per SAT review and INFUSE rocket flight) (Haughton et al (2024))
- ASIC electronics for 100 mm format XS at TRL3 (Curtis et al (2024)
- Open face 100 × 100 mm$^2$ format XS detector with HWO goal spatial resolution and high 100 nm QE at TRL6+ (Haughton et al (2024))
- Timepix/medipix single/double detector and sealed tube demonstration (TRL 3/4) (Tremsin & Vallerga (2020))

## Multi-Object Selection

### Discussion

Multi-object spectroscopy (MOS) is a technology development priority of both the Decadal Survey: Pathways to Discovery in Astronomy and Astrophysics for the 2020s (PDAA) and the current NASA Cosmic Origins Program. Aperture control methods that are popular in ground based MOS applications e.g., robotically configured fibers and punch plates) are not practical options for spaceflight.

The main target for the MOS capability is the HWO Ultraviolet MOS spectrograph which is considered a primary instrument for the mission. The science drivers for the mission are being defined at this time by the new HWO Project Office (PO) and call for the ability to take multiple spectra in sparsely populated fields and across both extended objects and point like targets, ranging in context from cosmological targets, to regions of star formation across galaxies, to individual resolved extended objects to solar system targets. In all cases a wide field of view (several arcminutes) is required with individual apertures subtending a few hundred milli-arcseconds, while delivering contrast performance between adjacent spectral channels of a part in $10^5$ for stellar sources. The MOS assembly would need be located at an intermediate point in the optical path of a telescope system to enable this high contrast.

## Current State of the Art

Proposal partnership POC: Alexander Kutyrev, NASA Goddard, alexander.s.kutyrev@nasa.gov

*Microshutter Arrays (MSA)*

A MSA functions as a programmable field slit mask. The array can provide any pattern of slits corresponding to sparsely distributed sources on the sky (similar to a punch plate). Our first generation MSA technology enabled realization of the James Webb Space Telescope (JWST) Near Infrared Spectrometer (NIRSpec). This prior generation technology involved a combination of electrostatic and magnetic actuation that results in a heavy complex mechanical assembly that does not scale to larger formats required by the Habitable Worlds Observatory. The next generation MSA (NGMSA) technology eliminates the need for the magnetic actuation system used in the JWST MSA and is scalable to a larger field of view.

To deliver the larger field of view, even with the larger form factors envisioned for the NGMSA, 2-side buttability would still be required. The program is currently supported by the three year SAT (FY24 FY26) and GSFC IRAD programs.

The NGMSA system (Kutyrev 2023) consists of the microshutter array itself on a mechanical structure and the control drive electronics. The NGMSA technology eliminates the macro-mechanisms needed by the previous JWST MSA magnetic actuation and is scalable for large-field-of-view MOS applications.

In 2022, the NASA PCOS/COR program office has vetted our current system TRL as follows. TRL array =4; TRL electronics = 3. The publicly available SAT report can be found at https://www.astrostrategictech.us/index.cfm. After thoroughly comparing the TRL definitions, hardware fidelity, and exit criteria, we confirm that the TRL assessment system utilized by the PCOS/COR Program Office aligns perfectly with that of the HWO Project Office. The NGMSA TRL assessment vetted by the PCOS/COR Program Office fully satisfies the TRL requirements set by the HWO Project Office.

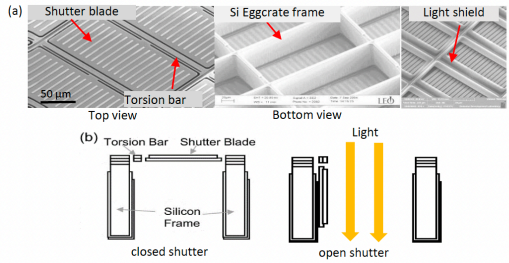
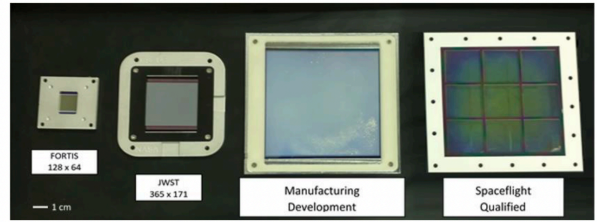

(a) Scanning electron microscopy (SEM) image of a microshutter array with component details. The light shield covers the gaps between the shutter blade, torsion bar and the silicon eggcrate frame. (b) When actuated, the shutter blade rotates and opens the shutter.

*MSA format development trend for Cosmic Origins is shown. Our pilot and JWST formats are shown on the left. A concept and manufacturing development array produced by our fiscal year FY18 Strategic Astrophysics Technology (SAT) project is shown in the center. Our FY18 SAT project also produced a ruggedized design that is suitable for spaceflight is shown on the right and is the subject of the current development. **The design shown on the right has been demonstrated through test to general environmental verification standard (GEVS) vibration levels for evolved expendable launch vehicle (EELV) launch**.*

NGMSA Sounding Rocket Flight Testbed: Off Axis FORTIS (PI: Stephan McCandliss/Johns Hopkins University). In collaboratation with the JHU team the 64x128 microshutter arrays have been flight tested as a multi-object selector mask for the OFF Axis Far UV Off Rowland circle Telescope for Imaging and Spectroscopy (Off Axis FORTIS). The NGMSA successfully operated in the FORTIS instrument during the 2019 and 2024 sounding rocket flights. The next flight is scheduled in 2026.

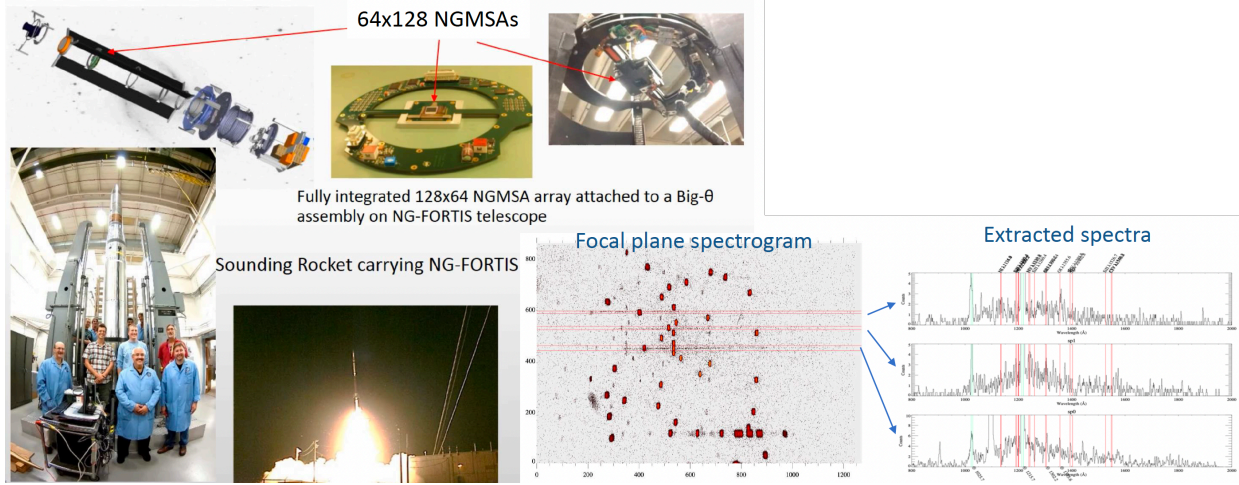

*Digital Micromirror Devices*

Proposal partnership POC: Massimo Roberto, STScI, robberto@stsci.edu

DMDs were first developed in the 1980s at Texas Instruments (TI) and have primarily been used for projection display applications (DLP technology, US Patent 5061049). A DMD consists of an array of millions of square micromirrors placed on top of the CMOS circuitry which allows the mirrors to be individually addressed and tilted into one of two stable states. During operation, the micromirrors are tilted at + and - 12 degrees from the device normal at frequencies of several kilohertz. The rapid binary modulation between the two states is used to create hundreds of digital pictures per second, i.e. a series of gray scale "photograms" that eventually create a color video when illuminated by a synchronized RGB light source. DMDs are a mature technology with impressive reliability and are generally regarded as the most successful MEMS technology ever developed.

MOS that use DMD as the reflective multiplexing mechanism utilize both positive and negative tilt states of the micromirrors as shown in Figure 2. In an MOS the focal plane is projected on the DMD. Each micromirror represents a potential slit reflecting light from selected target sources towards the spectrograph channel on the right, while the remaining micromirrors redirect the FOV towards an imaging channel. The imaging channel thus functions as a slit-viewing device and/or can be used to perform deep imaging in parallel with the spectroscopic data acquisition. The locations and sizes of slits are determined by the pattern deployed on the DMD. The slit mask (i.e. DMD pattern) can be reconfigured in a matter of seconds.

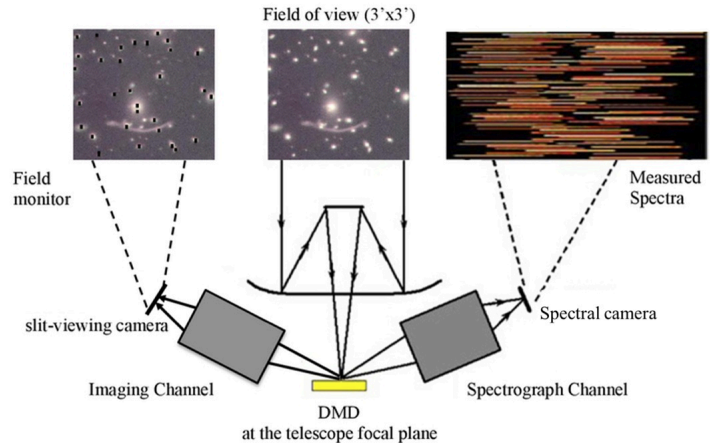

Figure 2: schematic illustration of a typical layout for a DMD-based MOS

Several DMD-based MOS have been deployed on ground-based telescopes including RITMOS at RIT and recently the NSF funded SAMOS at JHU (Robberto, 2016, Piotrowski, 2022, 2023, 2024). SAMOS has been recently deployed at the 4.1m SOAR telescope on Cerro Pachon, Chile to fully exploit the outstanding optical quality delivered by the Ground Layer Adaptive Optics system (SAM) over a 3 x 3 arcminutes field of view (Figure 2).

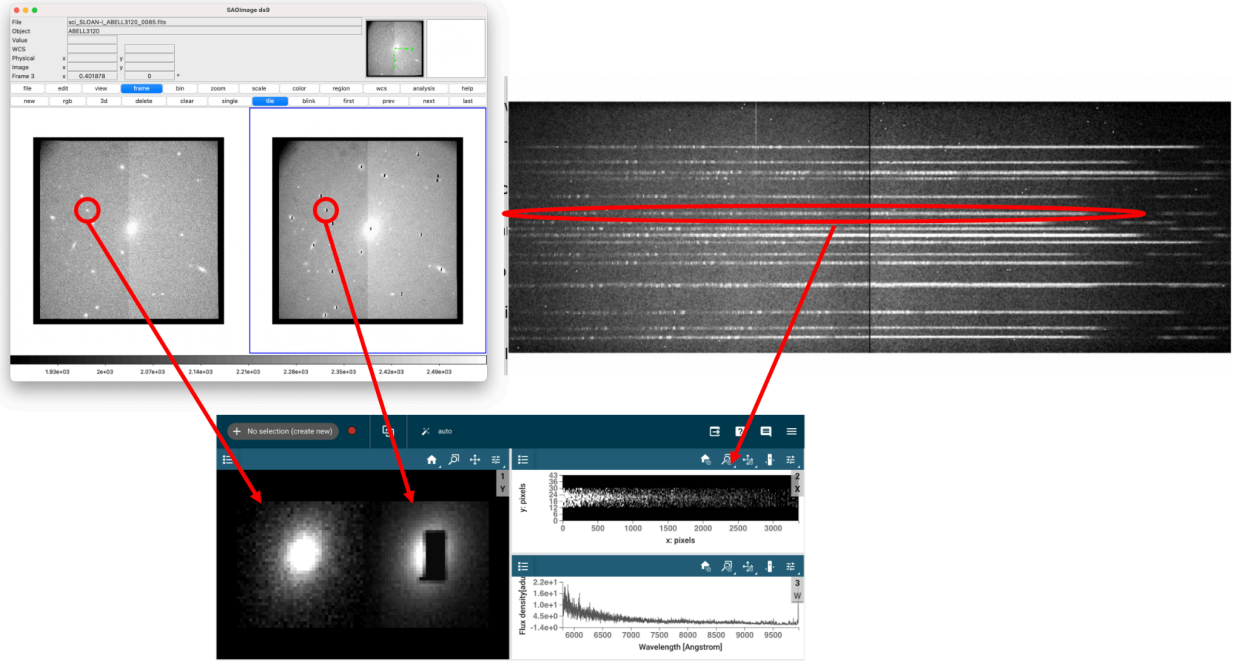

Figure 2: SAMOS data taken during the commissioning run in October 2024 for the galaxy cluster Abel 3120: top row, from left to right: the first target acquisition image on the imaging channel; the same field with the slits configured based on the pre-selected RADEC coordinates of the targets; the simultaneous MOS spectra, raw data, 900s integration. Bottom row: the same data for a random galaxy displayed after quick-look reduction using the JWST MOSVIZ Data Analysis Tool. Full moon conditions, no laser guide.

Commercially available DMDs have ~10x more slit elements with immaculate functionality and high fill factor, but are typically optimized for display applications that require fast pixel transition time and frame refreshing rate, as well as compact sizes. Several key design features for video display such as mirror size, tilt angle, and illumination of the DMD set the ultimate performance limitations of MOS-based on commercial DMDs. Before reviewing the main limitations of commercially available DMDs, we briefly summarize the main model of interest for astronomical applications:

- TI "UV": DLP9991UUV - These are devices that Texas Instrument has specifically designed for UV application. They come in the 920x1080 HDTV format with 10.8 μm mirror size. Their throughput (window coating) is "UV-optimized" for the 343-410 nm range.

## UV Gratings & Filters

### Discussion

*UV Gratings*

The dispersive elements needed in any UV spectrograph need particular performance metrics met to offset the higher scatter prevalent in the UV over, say, the visible and NIR bands. Mitigating scatter and improving both reflective throughput and spectral resolution can be achieved by more stringent manufacture methods that focus on increasing the accuracy of machining methods to produce cleaner and better defined grating structures. We present here a number of potential ways to do this to build on the current state of the art.

*UV Filters*

The construction of efficient UV filters has been a relatively recent development (the first Al/MgF2 Fabry-Perot filters, with remarkable efficiency, were developed in the late 1960s-early 1970s: Bates et al. 1966; Malherbe et al., 1970; Fairchild et al., 1973) with the construction of Fabry-Perot like cavity interference filters with the appropriate choice and use of metal + dielectric materials of the appropriate thickness and uniformity to produce a well-defined passband that can be used to isolate the densely packed spectral diagnostic lines common in the UV. The advent of techniques such as ALD and new plasma vapor deposition techniques have now made this a possibility. Scientifically this allows HWO to perform UV imaging through the UVI (or even the HRI) at the native wavelength as opposed to using a filter technology that upconverts UV photons into visible band photons as was done on the Hubble Space Telescope, with all the losses that came along with that approach.

## Current State of the Art

### UV Gratings

Proposal partnership POC: Randy McEntaffer, Penn State Univ., rlm90@psu.edu

### Echelle gratings

Researchers have created echelles with high blaze angles that have demonstrated high diffraction efficiency and low scatter (Kruczek et al (2022))

### Aberration-correcting UV gratings

Researchers have established a preliminary process to pattern curved grooves on curved substrates (Grise et al (2021); France et al (2022); Beasley et al (2019)). Researchers have established a range of angles that the groove can have relative to the Si-crystal structure to optimize the groove facet and minimize scatter (Grise et al (2021); France et al (2022)). Alternative blazing techniques are being developed to overcome chemical etching limitations (Miles et al. (2025); McCoy et al. (2020)).

### Gratings coating optimization

Conventional ion-etched blazed holographic gratings have been coated with advanced coatings for SPRITE/INFUSE programs (Kruczek et al (2022); France et al (2022))
- Post-coating, the gratings functioned to the requirements of those programs, however it is unclear whether they met HWO requirements for post-coating efficiency, scatter
- It is known that the HST-COS NUV gratings featured an interaction between the coating and grating that reduced efficiency
- Gratings should be designed with a model for coating dependencies to optimize performance for HWO – etched gratings offer the unique potential for fine groove shape design

### Electron-beam lithography UV gratings

Current work uses electron-beam lithography (EBL) to generate a grating pattern at virtually any groove density and groove layout, enabling high spectral resolving power. Using KOH etching, ion-beam etching (IBE), or thermally activated selective topography equilibration (TASTE), the EBL-defined groove pattern can be blazed to produce highly efficient gratings with low scatter properties. An appropriate UV reflective coating is then deposited for the grating's use in the final application.

*UV Filters*

Proposal partnership POC: Juan Larruquert, CSIC Madrid, Spain, j.larruquert@csic.es

FUV Reflective Filters

Five out of the six FUV filters identified as LUVOIR filters (F110M, F140M, F160M, F180M, F150W) are under development as part of a rocket payload (Nell et al (2024)). At present, F110M and F140M are already TRL 5, F150W is TRL 4, and F160M and F180M are TRL 3. F110M, F140M, F160M, and F180M will achieve TRL 6 by mid-2027 with the first rocket flight (Rodriguez de Marcos 2018; Farr et al. 2024; 2025). The F120M filter called out in the LUVOIR Final Report remains to be investigated, designed, and demonstrated.

Extra-Narrowband FUV Reflective Filters

Coatings with bandwidths in the order of ~4.5-5 nm for filters tuned at FUV wavelength longer than 120 nm have been developed; they are based on a new material combination, $MgF_2$ and $AlF_3$ (Lopez-Reyes et al. 2024). This figure provides examples of extra-narrowband filters that have been prepared:

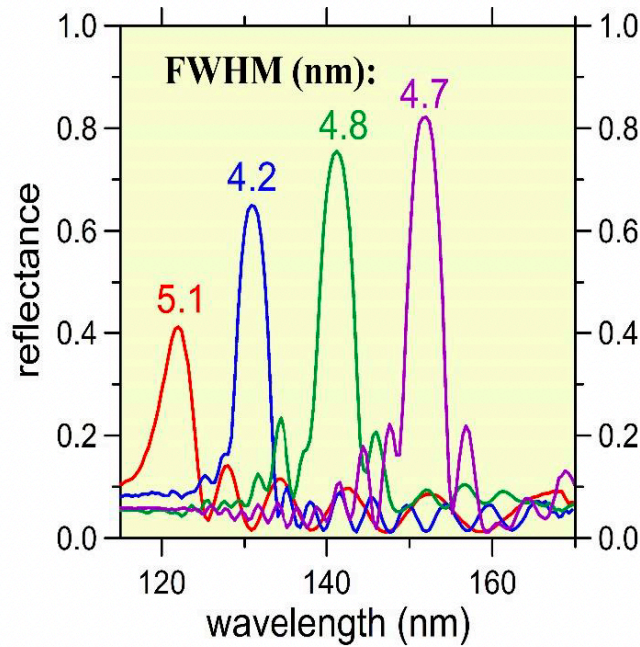

These extra-narrow bands have been only recently demonstrated and they may enable observations that were not possible before, which makes it adequate to elevate their TRL for their possible use in HWO. Here are promising spectral lines for novel imaging observations for astrophysics and planetary physics:

At present, extra-narrowband filters have been developed tuned at wavelengths between 120 and 185 nm (Larruquert et al 2022). A test of extra-narrowband mirrors under relevant environments (storage in 30-95% relative humidity for short periods, gamma radiation of 11 MRad, and thermal cycling in the -70-80ºC range) to increase TRL from 3 to 5 performed at GOLD-Instituto de Optica-CSIC will be reported soon.

Transmission Filters with Strong Rejection from the NUV to the NIR

Filters with transmittance in the visible below $10^{-6}$ have been fabricated for Chinese flight missions, reports about all these filters were provided to NSSC (China) and they are unpublished. The filters are installed in the IPM instrument on board of Chinese FY-3D (since 2017) and FY-3E (since 2021) satellites in a polar orbit.

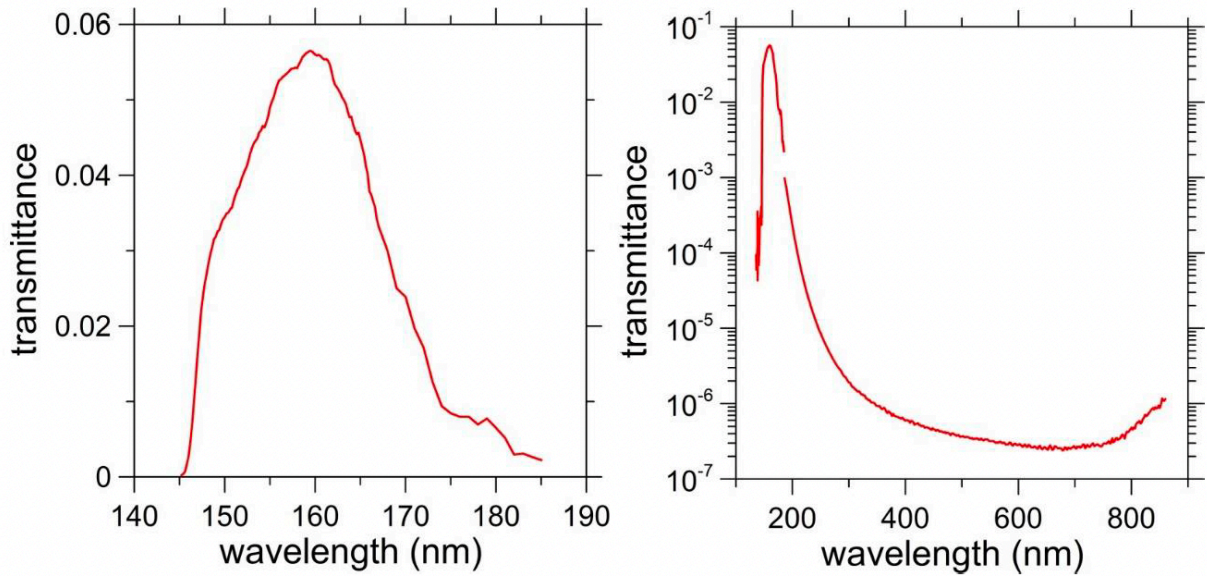

Such strong rejection of the visible may enable using detectors that are sensitive to the visible. This, along with the fact that an imager geometry is simplified with the use of transmittance filters, makes transmittance filters a valuable option.